# Is String Theory in Knots ?

## Philip E. Gibbs[1]

Abstract

It is sometimes said that there may be a unique algebraic theory independent of space-time topologies which underlies superstring and p-brane theories. In this paper, I construct some algebras using knot relations within the framework of event-symmetric string theory, and ask the question "Is string theory in knots?".

**Keywords**

quantum gravity, pregeometry, knots, braid group, string model

---

[1] e-mail to phil@galilee.eurocontrol.fr

# Introduction

After fifteen years of intensive research in superstring theory there is still a deep mystery surrounding its underlying nature. String theory has been found to take on a number of different forms in background space-times with different dimensions and topologies. Much to the delight of string theorists, many, if not all of these forms may be tied together through various forms of reduction and duality. This has led some theorists to speculate that string theory must have a unique formulation which is purely algebraic and independent of any space-time background (e.g. Schwarz 1995).

It is generally believed that string theory has a huge hidden symmetry which is recovered at very high energies (Gross 1988). If we could find a suitable description of string symmetry then we would begin to see its foundations. For mathematicians, classifying symmetries has been a priority occupation throughout the 20th century. Even so, there are large gaps in our understanding of infinite dimensional symmetry algebras and nothing is yet known to mathematics which can include all the supposed symmetries of string theory while at the same time unifying space-time symmetries with internal gauge symmetry.

In previous work I have argued for an event-symmetric approach to string theory (Gibbs 1995a; 1995b). This has led me to an interesting Lie-superalgebra which, I have suggested, may be at least part of string symmetry. I will recall its definition in the next section.

In the Canonical Approach to quantum gravity it is known that knots states are of some importance in the loop representation. Similarities between the loop representation and string theory may be more than superficial (Baez 1993). This observation is my inspiration to seek algebraic constructions for string theory which make use of ideas from knot theory (see e.g. Kauffman 1991).

## Discrete Superstring Symmetry

Let $E$ be a set of $N$ space time events and let $V = span(E)$ be the $N$ dimensional vector space spanned by those events. Then define $T = Tensor(V)$ to be the free associative algebra with unit generated over $V$. The components of $T$ form an infinite family of tensors over $V$ with one representative of each rank. Multiplication is given by tensor products.

$$\Phi = \{\varphi, \varphi_a, \varphi_{ab}, \varphi_{abc}, \ldots\}$$
$$\Phi^1 \Phi^2 = \{\varphi^1 \varphi^2, \varphi^1 \varphi^2_a + \varphi^1_a \varphi^2, \varphi^1 \varphi^2_{ab} + \varphi^1_a \varphi^2_b + \varphi^1_{ab} \varphi^2, \ldots\}$$

Here the indices *a, b, c* run over events.

The basis of this algebra already has a geometric interpretation as open strings passing through a sequence of events with arbitrary finite length. Multiplication of these strings consists merely of joining the end of the first to the start of the second. We can denote this as follows,

Algebraic String Theory

$$\Phi = \varphi + \sum_{a}\varphi_{a}a + \sum_{a,b}\varphi_{ab}ab + \sum_{a,b,c}\varphi_{abc}abc+\ldots$$

We now construct a new algebra by adding an extra connectivity structure to each string consisting of arrows joining events. There must be exactly one arrow going into each string and one leading out. This structure defines a permutation of the string events so there are exactly *K!* ways of adding such a structure to a string of length *K*.

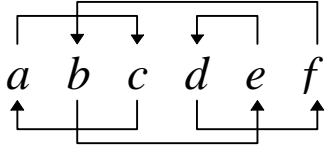

These objects now form the basis of a new algebra with associative multiplication consisting of joining the strings together as before, while preserving the connections. Finally the algebra is reduced modulo commutation relations between events in strings which are defined schematically as follows,

$$a\ b\ +\ b\ a\ =\ 2\delta_{ab}$$

These are partial relations which can be embedded into complete relations. Closed loops which include no events are identified with unity. For example, the lines can be joined to give,

$$a\ b\ +\ b\ a\ =\ 2\delta_{ab}$$

This example shows the cyclic relation on a loop of two events. The arrows can be joined differently to give another relation,

$$a\ b\ +\ b\ a\ =\ 2\delta_{ab}$$

which is the anti-commutation relation for loops of single events.

Algebraic String Theory

It is not immediately evident that the algebra thus defined does not reduce, through the relations above, to something quite trivial. To see that this does not happen we must at least verify that the relations are consistent with the symmetric group relations, especially the Yang Baxter equation. We proceed as follows,

$$a\ b\ c\ +\ a\ c\ b\ =\ 2\delta_{bc}\ a$$

$$a\ c\ b\ +\ c\ a\ b\ =\ 2\delta_{ca}\ b$$

$$c\ a\ b\ +\ c\ b\ a\ =\ 2\delta_{ab}\ c$$

$$c\ b\ a\ +\ b\ c\ a\ =\ 2\delta_{bc}\ a$$

$$b\ c\ a\ +\ b\ a\ c\ =\ 2\delta_{ac}\ b$$

$$b\ a\ c\ +\ a\ b\ c\ =\ 2\delta_{ab}\ c$$

Alternatively adding and subtracting these six equations gives zero on both sides. This ensures consistency of the relations.

By applying these relations repeatedly it is possible to reorder the events in any string so that the strings are separated into products of ordered cycles. Therefore we can define a more convenient notation in which an ordered cycle is indicated as follows,

$$(ab...c) \quad = \quad $$ 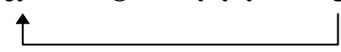

We can generate cyclic relations for loops of any length such as,

$$(ab) = -(ba) + 2\delta_{ab}$$
$$(abc) = (cab) + 2\delta_{bc}(a) - 2\delta_{ac}(b)$$
$$(abcd) = -(dabc) + 2\delta_{cd}(ab) - 2\delta_{bd}(d)(c) + 2\delta_{ad}(bc)$$

and graded commutation relations such as,

$$(a)(b) + (b)(a) = 2\delta_{ab}$$
$$(ab)(c) - (c)(ab) = 2\delta_{bc}(a) + 2\delta_{ac}(b)$$

Clearly the algebra has a $Z_2$ grading given by the parity of the length of string and it is therefore possible to construct an infinite dimensional Lie-superalgebra using the graded commutator.

The cycles of length one are the generators of a Clifford algebra and there is also a homomorphism from the full algebra onto a Clifford algebra defined by removing the loop structure from the strings. This is important from a physical point of view because of the roles that Clifford algebras play in physics. As well as being of fundamental importance in the construction of spinor representations of the rotation groups and their supersymmetry extensions, Clifford algebras are also isomorphic to the algebra of fermionic creation and annihilation operators. The algebra I have constructed here can be seen as the algebra of creation and annihilation operators for string states as well as an algebra of superstring symmetries describing strings formed from loops of fermionic partons in event-symmetric space-time.

## Knotted Algebras

The above discrete superstring symmetry is all very well except that strings are not made from discrete fermionic partons. They are defined as continuous loops, but at the same time they may be topological objects who's interactions can be determined by a finite number of discrete points. To try to capture this algebraically it may be necessary to envisage a string as being made from discrete partons with fractional statistics like anyons or parafermions. Such partons may be repeatedly subdivided into partons with smaller fractional statistics until a continuous limit is found. If strings are truly topological then only a finite sequence of subdivisions may be necessary to represent a particular interaction.

Motivated by these thoughts it is natural to seek some kind of deformation of the fermionic string algebra replacing the sign factors in the exchange relations with some general $q$-parameter. It is also natural to replace the loops which connect the partons with knots. In doing so we immediately hit upon a fortuitous coincidence. The construction of invariant knot polynomials makes use of Skein relations which are similar in to those we have already used. e.g.

$$q \;\; \times \;\; - \;\; q^{-1} \;\; \times \;\; = \;\; z \;\; || $$

This is the relation which defines the HOMFLY polynomial. Combining this with the algebra previously constructed suggests something like,

$$q \;\; a \;\; b \;\; - \;\; q^{-1} \;\; b \;\; a \;\; = z\delta_{ab}$$

The special case where there is only one event is related to the HOMFLY polynomial while the Lie superalgebra previously defined corresponds to the case $q = i$, $z = 2i$ with the sense in which two strings cross being disregarded. To completely define an algebra these relations can be embedded in knotted links.

It is not really clear if this is the best generalisation, an alternative is to make an algebra based in anyonic creation and annihilation operators $a, a^*$. The knot relations then specify the anticommutators for these operators when connected in a loop of string.

## Knotted Graphs

A further generalisation of knots which may be useful in algebraic string theory is knotted graphs. A knotted graph is like a knot except that multivalent nodes are permitted. To keep things simple we will allow just trivalent nodes subject to the relation,

$$\gtrless\!\!\lessgtr \;\; = \;\; \times$$

Just as knots can be orientated with arrows these networks can have an arrow and a flow integer assigned to each segment such that reversing the arrow is equivalent to changing the sign of the flow integer and the sum of the flows meeting at any node must be zero,

$$\xrightarrow{\quad n \quad} \;=\; \xleftarrow{\quad -n \quad}$$

Knotted networks are like knots in that it matters whether one string passes over or under another.

Let $\mathcal{A}$ be an associative algebra with unit over $\mathbb{C}$, and $g$ a grading of $\mathcal{A}$ over $\mathbb{Z}$. We can define a knotted algebra by placing elements of $\mathcal{A}$ on segments of a knotted network in such a way that the grading matches the flow. In addition the elements of the algebra are given an ordering across the page.

Multiplication in this system is then defined as the conjunction of the two networks with the strings of algebra elements concatenated together. Note that knotted networks don't have to be connected so this is already a well defined multiplication. To make it more interesting we must factor out the following relations,

$$q \;\; A \;\; B \;\; - \;\; q^{-1} \;\; B \;\; A \;\; = \;\; [A,B]_q$$

$q$ is a phase factor dependent on the grade of $A$ and $B$ and a fixed angle $\theta$.

$$q = e^{i\theta\, g(A)g(B)}$$

The q-commutator is defined as,

$$[A,B]_q = qA - q^{-1}B$$

The algebra thus defined also has a grading onto the integers defined by adding the grades of individual elements in a string. It is therefore possible to repeat these constructions to define higher dimensional structures. For example, if the base algebra $\mathcal{A}$ is an algebra of closed strings then in this way we can construct an algebra of branching surfaces which could be interpreted as string world sheets.

A further generalisation occurs when the base algebra is graded over $\mathbb{Z}_n$. It is then still possible to define the knotted graph algebra but only where $q$ is an $n^{th}$ root of -1. When $q$ is -1 or $i$ the associative algebra can be replaced by a Lie-algebra or Lie-superalgebra and in this case there is a homomorphism onto a smaller algebra in which the sense of knot crossings is ignored.

## Conclusions

In this breif introduction to algebraic string theory we have seen how it is possible to construct algebras using knot relations which can be interpreted as algebras of strings and higher dimensional objects in an event-symmetric space-time. The definitions given here leave many variations and generalisations possible. In later work I hope to explore which are the most interesting to string theory.

To complete the theory it will also be necessary to define the dynamics of the system and discover a correspondence with recognised string physics.